# Distributed order fractional sub-diffusion

Mark Naber [a]

Department of Mathematics

Monroe County Community College

Monroe, Michigan, 48161-9746

A distributed order fractional diffusion equation is considered. Distributed order derivatives are fractional derivatives that have been integrated over the order of the derivative within a given range. In this paper sub-diffusive cases are considered. That is, the order of the time derivative ranges from zero to one. The equation is solved for Dirichlet, Neumann, and Cauchy boundary conditions. The time dependence for each of the three cases is found to be a functional of the diffusion parameter. This functional is shown to have decay properties. Upper and lower bounds are computed for the functional. Examples are also worked out for comparative decay rates.

[a] Electronic mail: mnaber@monroeccc.edu

## 1. INTRODUCTION

Observations of physical systems that exhibit fractional diffusive behavior abound, see, e.g., Refs. 1, 2, 3, 4, 5, 6, and 7, and references therein. This process is also referred to as anomalous diffusion. The scale of fractional diffusive behavior ranges from the propagation of cosmic rays through the interstellar medium [3] to the diffusion of hydrogen atoms in solids [8]. This phenomena falls into two categories, sub-diffusion (also called ultra-slow) and super-diffusion (also called intermediate process). The former process is modeled using the fractional diffusion equation and the later is modeled using the fractional wave or fractional diffusion-wave equation. Physical systems that exhibit sub-diffusion diffuse at a rate much slower than that predicted by the usual (whole order derivative) diffusion equation. Likewise, super-diffusion is exhibited by systems that diffuse at a rate faster than for ordinary diffusion (hence the term anomalous diffusion to characterize diffusive systems that do not fit within the predictive range of the usual diffusion equation). An overview of fractional diffusive problems is given in Ref. 6. In this paper sub-diffusive cases will be treated. Super-diffusive problems will be considered in a later paper.

Time fractional diffusion and wave equations have been derived by considering continuous time random walk problems (CTRW), which are in general non-Markovian processes (see, e.g., Refs. 5, 9, 10, 11, 12, 13, and 14), and via diffusion in fractal media (see, e.g., Refs. 1, 2, 15, and 16). Fractional diffusion equations with the fractional derivative on the spatial derivative are used for studying Markovian processes. Nigmatulin was the first to derive a fractional diffusion equation by considering diffusion in a fractal media [16]. Fractal geometry may be thought of as a special case of porous media.  The physical interpretation of the fractional derivative in both cases is that it represents a degree of memory in the diffusing material.  In fractal media this may be viewed as arising from the spatial constraints imposed by the fractal structures.

In this paper the fractional diffusion equation is modified by integrating over all possible orders of the fractional time derivative.  This is called a distributed order derivative [9].  A physical interpretation would be that it is a time derivative acting on multiple time scales.  One possible application of this would be diffusion in a medium in which there is no fixed scaling exponent.  For example a multi-fractal medium or a medium in which there are memory effects over multiple time scales.

The fractional time derivative in fractional diffusion equations is either the Riemann-Liouville or the Caputo fractional derivative.  Each uses Riemann-Liouville fractional integration and derivatives of whole order.  The difference between the two fractional derivatives arises in the order of evaluation.  Riemann-Liouville fractional integration of order $\mu$ is defined as,

$$\mathrm{I}^{\mu}(f(t)) = \frac{1}{\Gamma(\mu)} \int_0^t \frac{f(\tau)\mathrm{d}\tau}{(t-\tau)^{1-\mu}}. \tag{1}$$

The next two equations define Riemann-Liouville and Caputo fractional derivatives of order $v$, respectively,

$$^{RL}\mathrm{D}_t^v f(t) = \frac{\mathrm{d}^k}{\mathrm{d}t^k}\left(\mathrm{I}^{k-v} f(t)\right), \tag{2}$$

$$^C\mathrm{D}_t^v f(t) = \mathrm{I}^{k-v}\left(\frac{\mathrm{d}^k}{\mathrm{d}t^k} f(t)\right). \tag{3}$$

Where $k-1 \leq v < k$.  See Ref. 7 for a derivation and further discussion.  The Caputo fractional derivative first computes an ordinary derivative followed by a fractional integral to achieve the desired order of fractional derivative. The Riemann-Liouville fractional derivative is computed in the reverse order.

The desire to formulate initial value problems for physical systems leads to the use of Caputo fractional derivatives rather than Riemann-Liouville fractional derivatives. Consider the Laplace transform of the Riemann-Liouville fractional derivative,

$$\mathcal{L}_t\left\{{}^{RL}\mathrm{D}_t^v f(t)\right\} = s^v F(s) - \sum_{k=0}^{n-1} s^k \left({}^{RL}\mathrm{D}_t^{v-k-1} f(t)\right)\Big|_{t=0}. \tag{4}$$

The initial conditions, $\left({}^{RL}\mathrm{D}_t^{v-k-1} f(t)\right)\Big|_{t=0}$ for $k = 0, \ldots, n-1$, are fractional order derivatives (see Ref. 17 for a detailed discussion of these objects). When studying a physical system, initial conditions are typically conditions that can be measured or imposed on the system. As yet there is no physical interpretation for $\left({}^{RL}\mathrm{D}_t^{v-k-1} f(t)\right)\Big|_{t=0}$ (see Ref. 18 for a fractal interpretation of the fractional order integral). Some authors using equations with Riemann-Liouville fractional derivatives to model physical systems have added terms to the diffusion equation to eliminate these unphysical terms (see, e.g., Ref. 10).

The Laplace transform of the Caputo fractional derivative is

$$\mathcal{L}_t\left\{{}^{C}\mathrm{D}_t^v f(t)\right\} = s^v \mathrm{F}(s) - \sum_{k=0}^{n-1} s^{v-k-1} \left(\mathrm{D}_t^k f(t)\right)\Big|_{t=0}. \tag{5}$$

In this case the initial conditions are well understood from a physical point of view. For example if $f(t)$ represents position then $\left(\mathrm{D}_t^0 f(t)\right)\Big|_{t=0}$ is the initial position, $\left(\mathrm{D}_t^1 f(t)\right)\Big|_{t=0}$ is the initial velocity, etc. (see Podlubny[7] section 2.4 for a more detailed discussion of this point). For the remainder of this paper, $\mathrm{D}_t^v$ shall denote the Caputo fractional derivative of order $v$ with $0 < v < 1$. It is assumed that the reader has a familiarity with fractional calculus.

In Sec. 2 the distributed order fractional diffusion equation is defined for the sub-diffusive case. In Secs. 3, 4, and 5, the Dirichlet, Neumann, and Cauchy problems are solved. It is shown that they all have a similar time dependence on a functional of the diffusion parameter. The diffusion parameter takes the place of the reciprocal of the diffusion constant and depends functionally on the order of the time derivative. Section 6 determines some boundedness properties of the functional of the diffusion parameter. Examples are also worked out for comparative decay rates. That is, given two diffusive systems with identical boundary and initial conditions, but different diffusion parameters, which will diffuse at a slower rate? Concluding remarks are given in Sec. 7.

## 2. DISTRIBUTED ORDER FRACTIONAL DIFFUSION EQUATION

1 + 1 dimensional fractional diffusion (for the non-Markovian case) is usually studied by considering the following equation (ignoring sources and boundary conditions),

$$D_t^\mu u(x,t) = c \frac{\partial^2}{\partial x^2} u(x,t). \tag{6}$$

$c$ is the diffusion coefficient, $0 < \mu < 1$ for sub-diffusion and $1 < \mu < 2$ for super-diffusion. It is important to note the sign of the diffusion coefficient. A positive diffusion coefficient indicates that the diffusing material is transported from a region of higher density to a region of lower density. If the diffusion coefficient were negative the diffusing material would be seen to accumulate rather than diffuse. There are several other fractional diffusion equations (see, e.g., Ref. 15 and references therein). An examination of the different types by Zeng and Li[15] indicates that Eq. (6) is the most physically relevant for the non-Markovian case in one space and one time dimension.

For distributed order sub-diffusion[9], the following equation is considered,

$$\int_0^1 C(v)\, D_t^v u(x,t)\, dv = \frac{\partial^2}{\partial x^2} u(x,t). \tag{7}$$

The reciprocal of the diffusion constant now depends on the order of the time derivative. This is an interesting mathematical generalization of Eq. (6) and which does have physical justification. Chechkin, et. al.[9] considered a particular form of $C(v)$ to study time evolution of a probability density function with a non-unique Hurst exponent. The Hurst exponent is a measure of the smoothness of a fractal time series.[22] This is a useful quantity for studying processes that have a long memory. For example, equation (7) should be useful for studying systems whose mean square displacement grows slower than that governed by a power law.[23] Possible applications of this would be diffusion in a medium in which there is no fixed scaling exponent. For example a multi-fractal medium or a medium in which there are memory effects over multiple time scales.

Equation (6) can be recovered by letting $C(v) = \frac{1}{c}\delta(v-\mu)$. $C(v) \geq 0$ and is not zero everywhere. This restriction is imposed for the physical reasons given above. The units of $C(v)$ are $(time)^v (length)^{-2}$. $C(v)$ will be referred to as the diffusion parameter. Due to the integral over the order of the time derivative a solution by similarity transformation is not possible (see, e.g., Refs. 18 and 19).

## 3. THE DIRICHLET PROBLEM

As a first case, consider Eq. (7) with the following initial and boundary conditions;

$$u(x,0) = f(x),$$
$$u(0,0) = u(1,0) = 0. \tag{8}$$

$f(x)$ is the initial distribution of the diffusive material (or the initial temperature). The spatial boundary conditions indicate that the diffusive material is removed at the endpoints, or if temperature is being considered, the endpoints are held at a fixed temperature. Separation of variables can be used to solve the equation. Let $u(x,t) = A(x)B(t)$, then Eq. (7) can be written as

$$\frac{1}{B(t)} \int_0^1 C(v) D_t^v B(t) \, dv = \frac{1}{A} \frac{d^2 A(x)}{dx^2} = -\lambda. \tag{9}$$

$\lambda$ is a real constant (the eigenvalue for the equation). The spatial component of the equation is solved using a Fourier series.

$$\lambda = n^2 \pi^2 \tag{10}$$

$$A(x) = \sum_{n=1}^{\infty} a_n \sin(n\pi x) \tag{11}$$

The $a_n$ are the Fourier coefficients that give $f(x)$.

The time dependent differential equation is solvable by Laplace transform. Let $b_n(s)$ denote the Laplace transform of $B_n(t)$ (i.e. $\mathcal{L}_t\{B_n(t)\} = b_n(s)$). The subscript $n$ is now included to indicate which eigenvalue is considered. Note that $B_n(0) = 1$ is used in the transform,

$$\mathcal{L}_t\left\{\int_0^1 C(v) D_t^v B_n(t) \, dv + (n\pi)^2 B_n(t) = 0\right\}, \tag{12}$$

$$\int_0^1 C(v)\left\{s^v b_n(s) - s^{v-1}\right\}dv + (n\pi)^2 b_n(s) = 0. \tag{13}$$

Isolating $b_n(s)$ yields

$$b_n(s) = \frac{\int_0^1 C(v)s^{v-1}dv}{\int_0^1 C(\mu)s^\mu d\mu + (n\pi)^2}. \tag{14}$$

To find the inverse Laplace transform for $b_n(s)$, the inverse transform is worked out for the reduced problem

$$b_n(v,s) = \frac{s^{v-1}}{\int_0^1 C(\mu)s^\mu d\mu + (n\pi)^2}. \tag{15}$$

The inverse transform is given by the complex inversion formula

$$B_n(v,t) = \frac{1}{2\pi i}\int_{\gamma-i\infty}^{\gamma+i\infty} e^{st} b_n(v,s)ds. \tag{16}$$

The solution for the original differential equation is then given by

$$B_n(t) = \int_0^1 B_n(v,t)\, C(v)dv. \tag{17}$$

Equation (16) is the complex inversion formula for Laplace transforms. $s$ is taken to be a complex variable, $s = r e^{i\theta}$. $\gamma$ is an arbitrary real number chosen so that it lies to the right of all poles and branch points in the integral. This integral has a branch point at $s = 0$ due to $s^{v-1}$ in the numerator and $s^\mu$ in the denominator. Any poles will be simple and are solutions of

$$\int_0^1 C(\mu)s^\mu d\mu + (n\pi)^2 = 0. \tag{18}$$

The left hand side of Eq. (18) is an analytic function in any region not containing the origin. Consequently, any poles will be isolated.

*Theorem*: The only solutions to Eq. (18) lie on the negative real axis.

*Proof*: Recall that $C(\mu) \geq 0$ for all values of $\mu$ and is not zero everywhere. Suppose that there is a solution in the upper half plane, $s = r e^{i\theta}$, $0 < \theta < \pi$. Consider the imaginary part of Eq. (18)

$$\int_0^1 C(\mu) \, r^\mu \sin(\mu\theta) \, d\mu = 0. \tag{19}$$

Since $0 \leq \mu \leq 1$, then, $0 \leq \mu\theta \leq \pi$ is also true. This implies $\sin(\mu\theta) \geq 0$, in fact $\sin(\mu\theta) = 0$ only at the upper and lower boundaries of the integral. Thus, at all points over which the integral is evaluated the integrand is either positive or zero. Therefore, there are no values of $\theta \in (0, \pi)$ that will make Eq. (19) true. A similar argument shows that there are no solutions in the lower half plane $-\pi < \theta < 0$. If $\theta = 0$, the imaginary part of the equation is satisfied, however, the real part of the equation is not. At $\theta = 0$, the real part of the integral takes the form of,

$$\int_0^1 C(\mu) \, r^\mu \, d\mu + (n\pi)^2 = 0. \tag{20}$$

The integral will always yield a positive number and $(n\pi)^2$ is also always positive. Therefore, there are no solutions for $\theta = 0$. This leaves only the negative real axis for any possible solutions, which completes the proof.

The inverse Laplace transform can be evaluated using residues. Due to the branch point at the origin the usual Bromwich contour cannot be used. A branch cut along the negative Real($s$) axis must be made. That is, a cut from $-\infty$ into and then around the origin in a clockwise sense and then back out to $-\infty$. The usual Bromwich contour is continued after the cut. This is referred to as a Hankel contour.

$$\frac{1}{2\pi i} \oint_{Ha} \frac{e^{st} s^{\nu-1}}{\int_0^1 C(\mu) s^\mu d\mu + (n\pi)^2} \, ds = \text{Residue of} \left( \frac{e^{st} s^{\nu-1}}{\int_0^1 C(\mu) s^\mu d\mu + (n\pi)^2} \right) \tag{21}$$

Breaking the Hankel contour up into its individual pieces and isolating the part that gives the inverse Laplace transform yields

$$B_n(v,t) = \text{Residue of} \left( \frac{e^{st} s^{v-1}}{\int_0^1 C(\mu) s^\mu d\mu + (n\pi)^2} \right) - \frac{1}{2\pi i} \left\{ \int_{\gamma+i\infty}^{-\infty} e^{st} b_n(v,s) ds \right.$$
$$\left. + \int_{-\infty}^0 e^{st} b_n(v,s) ds + \lim_{r \to 0} \oint e^{st} b_n(v,s) is d\theta + \int_0^\infty e^{st} b_n(v,s) ds + \int_{-\infty}^{\gamma-i\infty} e^{st} b_n(v,s) ds \right\}. \quad (22)$$

Recall that the only possible poles are on the negative real axis and are thus excluded from the residue computation. It can be shown that the contributions along the arcs $\gamma + i\infty \to -\infty$ and $-\infty \to \gamma - i\infty$ go to zero as the radius of the arc goes to $\infty$. The contribution from the arc around the origin, due to the branch cut, also goes to zero as the radius of the arc goes to zero. This leaves the contribution along the path $-\infty \to 0$ and the contribution along the path $0 \to -\infty$. These two contributions give

$$B_n(v,t) = \frac{1}{2\pi i} \left\{ \int_0^\infty \left( \frac{e^{-rt} r^{v-1} e^{i\pi v} dr}{\int_0^1 C(\mu) r^\mu e^{i\pi\mu} d\mu + (n\pi)^2} \right) - \int_0^\infty \left( \frac{e^{-rt} r^{v-1} e^{-i\pi v} dr}{\int_0^1 C(\mu) r^\mu e^{-i\pi\mu} d\mu + (n\pi)^2} \right) \right\}. \quad (23)$$

Notice that the second integral is the complex conjugate of the first integral. For that reason, only the imaginary part of the first integral is needed.

$$B_n(v,t) = \frac{1}{\pi} \text{Im} \left\{ \int_0^\infty \left( \frac{e^{-rt} r^{v-1} e^{i\pi v} dr}{\int_0^1 C(\mu) r^\mu e^{i\pi\mu} d\mu + (n\pi)^2} \right) \right\} \quad (24)$$

Eq. (24) is used with Eq. (17) to obtain $B_n(t)$. The resulting expression can be cast in a more useful form if the following functions are defined;

$$g(r,n) = \int_0^1 C(\mu) r^\mu \cos(\pi\mu) d\mu + (n\pi)^2, \quad (25)$$

$$h(r) = \int_0^1 C(\mu) r^\mu \sin(\pi\mu) d\mu. \quad (26)$$

Then $B_n(t)$ can be expressed as

$$B_n(t) = \frac{1}{\pi} \int_0^\infty \frac{e^{-rt}}{r} \frac{(n\pi)^2 h(r) \, dr}{g(r,n)^2 + h(r)^2}. \tag{27}$$

The final expression for $B_n(t)$ can be simplified to give

$$B_n(t) = \frac{1}{\pi} \int_0^\infty \frac{e^{-rt}}{r} \frac{(n\pi)^2 h(r)}{g(r,n)^2 + h(r)^2} dr. \tag{28}$$

$h(r) > 0$ for all $r$ values, hence, $B_n(t) > 0$ for all $n$ and $t$ values. Also notice that $B_n(t) \to 0$ as $t \to \infty$. $B_n(0) = 1$ is assured by the initial conditions used in the Laplace transform. The solution to the original diffusion problem can now be written as,

$$u(x,t) = \sum_{n=1}^\infty a_n B_n(t) \sin(n\pi x). \tag{29}$$

Note that $u(x,t) \to 0$ as $t \to \infty$. This is the expected physical behavior for a diffusion problem with the given boundary conditions.

## 4. THE NEUMANN PROBLEM

Now consider Eq. (7) with the following boundary conditions

$$\begin{aligned} u(x,0) &= f(x), \\ \frac{\partial}{\partial x} u(x,t)\bigg|_{x=0} &= \frac{\partial}{\partial x} u(x,t)\bigg|_{x=1} = 0. \end{aligned} \tag{30}$$

The spatial boundary conditions indicate that the material cannot pass through the endpoints. For the temperature case this is a rod with insulated ends. This equation can also be solved by separation of variables. Let $u(x,t) = A(x)B(t)$ just as before. In this case, the sine functions are replaced by cosine functions and $n$ can now be zero.

$$A(x) = \sum_{n=0}^\infty a_n \cos(n\pi x) \tag{31}$$

The Laplace transform of the time dependent differential equation is written as before and solved in the same fashion,

$$b_n(s) = \frac{\int_0^1 C(v) s^{v-1} dv}{\int_0^1 C(\mu) s^{\mu} d\mu + (n\pi)^2}. \tag{32}$$

For $n$ not being zero the inverse Laplace transform is worked out the same way as for the previous problem. If $n$ is zero Eq. (32) becomes

$$b_0(s) = \frac{\int_0^1 C(v) s^{v-1} dv}{\int_0^1 C(\mu) s^{\mu} d\mu} = \frac{1}{s}, \tag{33}$$

and the inverse Laplace transform is

$$B_0(t) = 1. \tag{34}$$

The solution to the original problem is then

$$u(x,t) = \sum_{n=0}^{\infty} a_n B_n(t) \cos(n\pi x). \tag{35}$$

Just as in the non-fractional case, $u(x,t) \to a_0$ as $t \to \infty$.

## 5. CAUCHY PROBLEM

Now consider Eq. (7) on an infinite spatial domain.

$$u(x,0) = f(x), \quad |u(x,t)| < M, \quad -\infty < x < \infty, \quad t > 0 \tag{36}$$

Taking the Fourier transform of Eq. (7) with respect to $x$ gives

$$\int_0^1 C(v) \, D_t^v U(\lambda,t) dv = -\lambda^2 U(\lambda,t). \tag{37}$$

Where

$$U(\lambda,t) = \mathcal{F}_x(u(x,t)) = \int_{-\infty}^{\infty} u(x,t) e^{-i\lambda t} dt \tag{38}$$

is the spatial Fourier transform of u. The Laplace transform is now taken with respect to the time variable. Denote $\mathcal{L}_t\{U(\lambda,t)\} = \tilde{U}(\lambda,s)$, and Eq. (37) becomes

$$\int_0^1 C(v)\left\{s^v \tilde{U}(\lambda,s) - s^{v-1} U(\lambda,0)\right\}dv + \lambda^2 \tilde{U}(\lambda,s) = 0. \tag{39}$$

Isolating $\tilde{U}(\lambda,s)$ gives

$$\tilde{U}(\lambda,s) = \frac{U(\lambda,0)\int_0^1 C(v) s^{v-1} dv}{\int_0^1 C(\mu) s^\mu d\mu + \lambda^2}. \tag{40}$$

The inversion of the Laplace transform is carried out just as is done for the other two cases,

$$U(\lambda,t) = \frac{U(\lambda,0)}{\pi}\int_0^\infty \left(\frac{e^{-rt}}{r}\frac{\lambda^2\,\mathrm{h}(r)}{\mathrm{g}(r,\lambda/\pi)^2 + \mathrm{h}(r)^2}\right)dr. \tag{41}$$

$\lambda/\pi$ now replaces $n$. This leaves undoing the Fourier transform and identifying the initial condition,

$$U(\lambda,0) = \mathcal{F}_x(f(x)) = F(\lambda), \tag{42}$$

$$U(\lambda,t) = \frac{F(\lambda)\lambda^2}{\pi}\int_0^\infty \frac{e^{-rt}\mathrm{h}(r)dr}{r(\mathrm{g}(r,\lambda/\pi)^2 + \mathrm{h}(r)^2)}. \tag{43}$$

The inverse transform may be written as a convolution integral. Define the following function,

$$G(x,t) = \mathcal{F}_\lambda^{-1}\left\{\int_0^\infty \frac{e^{-rt}\mathrm{h}(r)dr}{r(\mathrm{g}(r,\lambda/\pi)^2 + \mathrm{h}(r)^2)}\right\}. \tag{44}$$

Note that the function within the integral of Eq. (44) is even in $\lambda$, so,

$$G(x,t) = \frac{1}{\pi}\int_0^\infty\int_0^\infty \frac{e^{-rt}\mathrm{h}(r)}{r(\mathrm{g}(r,\lambda/\pi)^2 + \mathrm{h}(r)^2)}\cos(\lambda x)drd\lambda. \tag{45}$$

The final solution is then given by

$$u(x,t) = -f(x) * \frac{\partial^2}{\partial x^2} G(x,t), \tag{46}$$

where $*$ denotes the convolution product. Note that $-\partial_x^2 G(x,t)$ is a positive quantity that goes to zero as $t$ goes to infinity.

## 6. PROPERTIES OF THE SOLUTIONS

Solutions to all three problems depend on the same quantity,

$$\int_0^\infty \frac{e^{-rt}}{r} \frac{h(r)}{g(r,n)^2 + h(r)^2} dr. \tag{47}$$

In the Cauchy problem replace $n$ with $\lambda/\pi$. An understanding of the above integral is important to an understanding of solutions to distributed order fractional diffusion problems. It has already been noted that the integral goes to zero as $t$ goes to infinity, and that the integral is equal to $1/n^2\pi$ when $t$ is zero. In what follows, upper and lower bounds to the integral will be determined. Due to the $e^{-rt}$ term in Eq. (47), the integral may be viewed as a Laplace transform.

An upper bound for the integral can be determined by finding an upper bound for $h(r)$ and a lower bound for $g(r,n)^2 + h(r)^2$. The trick here is to find an upper bound that will still leave the integral bounded. An upper bound for $h(r)$ is given by

$$h(r) \leq \begin{cases} \dfrac{r - \ln(r)}{\sqrt{r}} \tilde{C} & 0 \leq r \leq 1 \\ rM & 1 \leq r < \infty \end{cases} \tag{48}$$

where, $M = \int_0^1 C(\mu)\sin(\pi\mu)d\mu \neq 0$ and $\tilde{C} = \int_0^1 C(\mu)d\mu \neq 0$. Now consider the quantity $g(r,n)^2 + h(r)^2$. Since $C(\mu)$ and $\sin(\pi\mu) \geq 0$ over the range of integration, the function $h(r)$ is a monotonically increasing function of $r$ whose minimum value is $h(0) = 0$. The same question is not so clear for $g(r,n)$ due to the cosine term,

$$g(r,n) = \int_0^1 C(\mu) r^\mu \cos(\pi\mu) d\mu + (n\pi)^2. \tag{49}$$

$g(r,n)$ can be either positive or negative, or have roots, depending on the properties of $C(\mu)$. However, $g(r,n)$ appears as $g(r,n)^2$ in the problem being considered. Since $g(r,n)$ is real, $g(r,n)^2 \geq 0$ for all r. Note that $g(0,n)^2 = (n\pi)^4$. The minimum value of $g(r,n)^2$ will be less than or equal to $(n\pi)^4$. This minimum value may be zero. Denote the location of this minimum by $r_0$, then, $g(r_0,n)^2 \leq (n\pi)^4$. Hence $g(r,n)^2 + h(r)^2 \leq m$ where $m = \min((n\pi)^4, g(r_0,n)^2 + h(r_0)^2)$. Note that $m \neq 0$. Combining the upper bound for $h(r)$ and the lower bound for $g(r,n)^2 + h(r)^2$ gives

$$\int_0^\infty \frac{e^{-rt}}{r} \frac{h(r)}{g(r,n)^2 + h(r)^2} dr \leq \frac{\tilde{C}}{m} \int_0^1 e^{-rt} \frac{r - \ln(r)}{\sqrt{r}} dr + \frac{M}{m} \int_1^\infty e^{-rt} dr. \tag{50}$$

Notice that both integrals go to zero monotonically in time. The first integral on the right hand side of Eq. (50) can be evaluated using a power series and the second integral can be evaluated directly. The final inequality is then

$$\int_0^\infty \frac{e^{-rt}}{r} \frac{h(r)}{g(r,n)^2 + h(r)^2} dr \leq \frac{M}{m\,t} + \frac{\tilde{C}}{m} \sum_{k=0}^\infty \frac{(-t)^k (k^2 + 2k + 7/4)}{k!(k+3/2)(k+1/2)^2}. \tag{51}$$

Now consider a lower bound for the integral in Eq. (47). To determine how this integral is bounded from below it must be understood how $h(r)$ is bounded from below and how $g(r,n)^2 + h(r)^2$ is bounded from above. For $0 \leq \mu \leq 1$ and $0 \leq r < \infty$ the following is true,

$$r^\mu \geq re^{-r}. \tag{52}$$

Using this inequality in the definition of $h(r)$ gives

$$h(r) \geq re^{-r}\int_0^1 C(\mu)\sin(\pi\mu)d\mu = re^{-r}M. \tag{53}$$

Using a trigonometric identity $g(r,n)^2 + h(r)^2$ can be written as

$$\int_0^1\int_0^1 C(\mu)C(\nu)r^{\mu+\nu}\cos(\pi(\nu-\mu))d\nu d\mu + 2(n\pi)^2\int_0^1 C(\mu)r^\mu\cos(\pi\mu)d\mu + (n\pi)^4. \tag{54}$$

Hence,

$$g(r,n)^2 + h(r)^2 \leq \int_0^1\int_0^1 C(\mu)C(\nu)r^{\mu+\nu}d\nu d\mu + 2(n\pi)^2\int_0^1 C(\mu)r^\mu d\mu + (n\pi)^4. \tag{55}$$

The above inequality can be further reduced using the Holder inequality for integrals (see Ref. 21). For example,

$$\int_0^1 C(\mu)r^\mu d\mu \leq \left(\int_0^1 C(\mu)^2 d\mu\right)^{1/2}\left(\int_0^1 r^{2\mu}d\mu\right)^{1/2} = \hat{C}\sqrt{\frac{r^2-1}{2\ln(r)}}, \tag{56}$$

where $\hat{C} = \sqrt{\int_0^1 C(\mu)^2 d\mu}$. This gives

$$g(r,n)^2 + h(r)^2 \leq \hat{C}^2 \frac{r^2-1}{2\ln(r)} + 2\hat{C}(n\pi)^2\sqrt{\frac{r^2-1}{2\ln(r)}} + (n\pi)^4. \tag{57}$$

This inequality can be simplified by noting $\frac{r^2-1}{2\ln(r)} \leq \frac{r^2+3}{3}$ and $\sqrt{\frac{r^2-1}{2\ln(r)}} \leq \frac{r+1}{2}$ then,

$$g(r,n)^2 + h(r)^2 \leq \hat{C}^2\frac{r^2+3}{3} + 2\hat{C}(n\pi)^2\frac{r+1}{2} + (n\pi)^4. \tag{58}$$

Further reducing gives

$$g(r,n)^2 + h(r)^2 \leq \left(\frac{\hat{C}^2}{3} + \hat{C}(n\pi)^2\right)r^2 + \left(\hat{C} + (n\pi)^2\right)^2. \tag{59}$$

This final expression is in a form that will allow the integral in Eq. (47) to be evaluated. The lower bound is given by

$$\int_0^\infty \frac{e^{-rt}h(r)dr}{r(g(r,n)^2 + h(r)^2)} \geq \frac{M}{\Omega\omega}\left\{\cos\left(\frac{\omega}{\Omega}(t+1)\right)\left(\frac{\pi}{2} - \text{Si}\left(\frac{\omega}{\Omega}(t+1)\right)\right) - \sin\left(\frac{\omega}{\Omega}(t+1)\right)\text{Ci}\left(\frac{\omega}{\Omega}(t+1)\right)\right\} \tag{60}$$

where, $\Omega = \sqrt{\left(\dfrac{\hat{C}^2}{3} + \hat{C}(n\pi)^2\right)}$ and $\omega = \sqrt{\hat{C} + (n\pi)^2}$, and Si and Ci are the sine and cosine integral functions. Please note that the sign convention is that of Ref. 21. The sign convention of Ref. 21 is the opposite of the convention in MAPLE. Despite the appearance of the trigonometric functions in Eq. (60), the lower bound decays monotonically in time, as one would expect for a diffusion problem.

An interesting physical question is that of comparative decay rates. Consider two diffusion processes with the same set of boundary and initial conditions but with two different decay parameters, say $C_1(\nu)$ and $C_2(\nu)$. The question now arises, which decay parameter will allow for slower diffusion? That is, what is the criterion for the following inequality to be true for all $t \geq 0$,

$$\int_0^\infty \frac{e^{-rt}}{r} \frac{h_1(r)}{g_1(r,n)^2 + h_1(r)^2} dr > \int_0^\infty \frac{e^{-rt}}{r} \frac{h_2(r)}{g_2(r,n)^2 + h_2(r)^2} dr, \tag{61}$$

where,

$$h_1(r) = \int_0^1 C_1(\nu) r^\nu \sin(\pi\nu) d\nu, \tag{62}$$

$$h_2(r) = \int_0^1 C_2(\nu) r^\nu \sin(\pi\nu) d\nu, \tag{63}$$

etc.

If Eq. (61) is true, then the diffusion process modeled by $C_1(\nu)$ will occur at a slower rate than that modeled by $C_2(\nu)$. A sufficient condition for Eq. (61) to be true is that,

$$\frac{h_1(r)}{g_1(r,n)^2 + h_1(r)^2} > \frac{h_2(r)}{g_2(r,n)^2 + h_2(r)^2} \tag{64}$$

for all $r \geq 0$. Though, the above constraint is clearly not necessary.

In each of the following examples the diffusion parameter is normalized to unity.

$$\int_0^1 C(v)\,dv = 1 \tag{65}$$

The following special cases are easily seen to be true. Let

$$C_1(v) = \delta(v - v_1) \tag{66}$$

and,

$$C_2(v) = \delta(v - v_2) \tag{67}$$

with $v_2 > v_1$, then Eq. (61) is true. The diffusion process with the lower order of time derivative will diffuse at a slower rate.

As another example, let

$$C_1(v) = c\delta(v - v_1) + (1-c)\delta(v - v_2), \tag{68}$$

and

$$C_2(v) = \delta(v - v_2), \tag{69}$$

where $0 \le c \le 1$. Then, Eq. (61) is again true. The diffusion process with lowest order of time derivative will diffuse slowest.

As another example consider the Dirichlet problem with an initial distribution of (see Eq. (11))

$$A(x) = \sin(\pi x). \tag{70}$$

In this case, $n = 1$ is the only non-zero term in eq. (11). Let the two decay parameters be given by

$$C_1 = \delta(v - v_1) \tag{71}$$

$$C_2 = \frac{1}{v_2 - v_1}\left(H(v - v_1) - H(v - v_2)\right) \tag{72}$$

where H is the Heaviside unit step function and $0 < v_1 < v_2 \le 1$. In this case

$$h_1 = r^{v_1} \sin(v_1 \pi), \tag{73}$$

$$g_1 = r^{v_1}\cos(v_1\pi) + \pi^2,\tag{74}$$

$$h_2 = \frac{\ln(r)\bigl(r^{v_2}\sin(v_2\pi) - r^{v_1}\sin(v_1\pi)\bigr) - \pi\bigl(r^{v_2}\cos(v_2\pi) - r^{v_1}\cos(v_1\pi)\bigr)}{(v_2 - v_1)\bigl(\ln(r)^2 + \pi^2\bigr)},\tag{75}$$

$$g_2 = \frac{\ln(r)\bigl(r^{v_2}\cos(v_2\pi) - r^{v_1}\cos(v_1\pi)\bigr) - \pi\bigl(r^{v_2}\sin(v_2\pi) - r^{v_1}\sin(v_1\pi)\bigr)}{(v_2 - v_1)\bigl(\ln(r)^2 + \pi^2\bigr)} + \pi^2.\tag{76}$$

Using Maple (or some similar computer algebra system) it is not difficult to show that $B(C_1) > B(C_2)$ for all t values (see Eq. (28)). This can be explained as follows; for $C_1$ all the diffusing material diffuses at a rate governed by the time derivative of order $v_1$. For $C_2$ the diffusing material can diffuse at rates governed by time derivatives of any order between $v_1$ and $v_2$ the rate of decay increasing as the order of the time derivative increases.

As a contrast to the previous example let

$$C_1 = \frac{1}{v_2 - v_1}\bigl(H(v - v_1) - H(v - v_2)\bigr),\tag{77}$$

$$C_2 = \delta(v - v_2).\tag{78}$$

In this case the diffusion modeled by $C_2$ will occur faster than that modeled by $C_1$. This is because the only decay channel available to the $C_2$ system allows for faster decay than the channels available to the $C_1$ system.

## 7. CONCLUSION

In this paper the problem of distributed order fractional sub-diffusion was considered. General solutions were found for each of the three boundary conditions considered. The spatial component of the solutions was found to be the same as in the non-fractional case. It should be noted, that since the fractional diffusion equation is separable, the solutions presented here could be adapted to multi-dimensional cases or even cases where the Laplacian is fractional[15]. The interesting part of the solutions was found to be the time dependent component. In all three cases the time dependence was described by the same functional expression (see Eq. 47). This expression can be viewed as a Laplace transform. Upper and lower bounds for the time component were computed and both were found to have decay-like shapes (monotonically decreasing functions of time).

An interesting open question raised in this paper is that of the necessary condition for Eq. (61) to be true. Four examples were presented, each with straightforward criteria and a physical interpretation. A resolution to the general question would allow for physical insights in the comparison between two diffusive processes.

**ACKNOWLEDGMENTS**

The author would like to thank Jean Krish, for several useful discussions and suggestions. The author would also like to thank P. Dorcey and S. Vincelli for helpful comments and a critical reading of the paper. The referee is also thanked for several suggestions and useful comments.

[1] H. E. Roman and M. Giona, "Fractional diffusion equations on Fractals: Three-dimensional case and scattering function," J. Phys. **A25**, 2107-2117 (1992).

[2] M. Giona and H. E. Roman, "Fractional diffusion equation for transport phenomena in random media," Physica A **185**, 82-97 (1992).

3) A. A. Lagutin and V. V. Uchaikin, "Fractional diffusion of cosmic rays," arXiv:astro-ph/0107230 v1 13 Jul (2001).

4) G. Erochenkova and R, Lima, "A fractional diffusion equation for a marker in porous media," Chaos **11**, 3 (2001).

5) J. Klafter, M. F. Shlesinger, and G. Zumofen, "Beyond Brownian motion," Physics Today, 33-39 Feb. (1996).

6) I. M. Sokolov, J. Klafter, and A. Blumen, "Fractional kinetics," **55**, No. 11, 48-54 Nov. (2002).

7) Podlubny, I.: *Fractional Differential Equations*. Academic Press, (1999).

8) R. Hempelmann, Hydrogen diffusion in proton conducting oxides and in nanocrystalline metals, in; Anomalous Diffusion From Basics to Applications Proceedings of the XIth Max Born Symposium, A. Pekalski and K. Sznajd-Weron Eds., Lecture Notes in Physics, vol. 519, Springer-Verlag (1999).

9) A. V. Chechkin, R. Gorenflo, and I. M. Sokolov, Retarding sub- and accelerating super-diffusion governed by distributed order fractional diffusion equations, arXive:cond-mat/0202213 v1 13 Feb. (2002).

10) B. I. Henry and S. L. Wearne, Fractional reaction-diffusion, Physica A, 276, 448-455 (2000).

11) E. Scalas, R. Gorenflo, F. Mainardi, and M. Raberto, Revisiting the derivation of the fractional diffusion equation, arXiv:cond-mat/0210166 v2 11 Oct. (2002).

12) A. V. Chechkin, V. Yu. Gonchar, and M. Szydlowski, "Fractional Kinetics for Relaxation and Superdiffusion in Magnetic Field," arXiv:physics/0107018 v1 9 Jul. (2001).

[13] R. Gorenflo and F. Mainardi, "Fractional calculus and stable probability distributions," Arch. Mech. **50** (3), 377-388 (1998).

[14] G. Rangarajan and M. Ding, First passage time problem for biased continuous-time random walks, Fractals **8** (2), 139-145 (2000).

[15] Q. Zeng and H. Li, Diffusion equation for disordered fractal media, Fractals **8** (1), 117-121 (2000).

[16] R. R. Nigmatullin, The realization of the generalized transfer equation in a medium with fractal geometry, Phys. Stat. Sol. B **133**, 425-430 (1986).

[17] R. Hilfer, "Fractional Diffusion based on Riemann-Liouville Fractional Derivatives," J. Phys. Chem B, **104**, 3914 (2000).

[18] R. Nigmatullin, The fractional integral and its physical interpretation, Theor. and Math. Phys. **90**, 242-251 (1992).

[19] E. Buckwar and Y. Luchko, "Invariance of a Partial Differential Equation of Fractional Order Under the Lie Group of Scaling Transformations," J. Math. Anal. and Ap. **227**, 81-97 (1998).

[20] F. Mainardi, "The fundamental solutions for the fractional diffusion-wave equation," Applied Mathematics Letters **9**, No 6, 23-28 (1996).

[21] M. R. Spiegel and J. Liu, Mathematical Handbook of Formulas and Tables, Schaum's Outline Series, McGraw-Hill (1999).

[22] M. Schroeder, Fractals, Chaos, Power Laws, W. H. Freeman and Company (1991).

[23] A.V. Chechkin, J. Klafter, and I.M. Sokolov, Fractional Fokker-Planck equation for ultraslow kinetics, http://arxiv.org/pdf/cond-mat/0301487.